\documentclass[aip,jcp,reprint,groupedaddress]{revtex4-1}

\usepackage{amsmath}
\usepackage{txfonts}

\usepackage[T1]{fontenc}
\usepackage{textcomp}

\usepackage{bm}
\bmdefine{\aVector}{a}
\bmdefine{\AVector}{A}
\bmdefine{\bVector}{b}
\bmdefine{\BVector}{B}
\bmdefine{\cVector}{c}
\bmdefine{\CVector}{C}
\bmdefine{\eVector}{e}
\bmdefine{\EVector}{E}
\bmdefine{\fVector}{f}
\bmdefine{\FVector}{F}
\bmdefine{\gVector}{g}
\bmdefine{\pVector}{p}
\bmdefine{\PVector}{P}
\bmdefine{\qVector}{q}
\bmdefine{\QVector}{Q}
\bmdefine{\rVector}{r}
\bmdefine{\RVector}{R}
\bmdefine{\sVector}{s}
\bmdefine{\uVector}{u}
\bmdefine{\vVector}{v}
\bmdefine{\VVector}{V}
\bmdefine{\muVector}{\mu}
\bmdefine{\OmegaVector}{\Omega}

\usepackage[pdftex]{graphicx}

\usepackage{enumerate}

\draft 

\begin{document}


\title{Two-dimensional percolation phenomena of single-component linear homopolymer brushes} 



\author{Yuki Norizoe}
\author{Hiroshi Jinnai}
\author{Atsushi Takahara}
\affiliation{Japan Science and Technology Agency (JST), Exploratory Research for Advanced Technology (ERATO), Takahara Soft Interfaces Project, Kyushu University, 744 Motooka, Nishi-ku, Fukuoka 819-0395, Japan}


\date{\today}

\begin{abstract}
Percolation phenomena of homopolymer brushes on a planar substrate are simulated using the molecular Monte Carlo method in 3 dimensions. The grafted polymers are isolated from each other at extremely low grafting density, whereas a continuous polymer layer covers the whole substrate when the density rises to extremely high values. This indicates that percolation clusters of the grafted polymers, bridging both the edges of the substrate, appear at an intermediate density. We construct phase diagrams of this percolation phenomenon. Critical phenomena at the transition are also studied.
\end{abstract}


\maketitle 



%
%

%

\section{Introduction}
\label{sec:Introduction}
Polymer brushes, \textit{i.e.} polymer chains grafted onto substrates, have been drawing broad attention of physicists, chemists, and others~\cite{Milner:1991,Soga:1995,O'Driscoll:2011,Guskova:2012,Wang:2009,Norizoe:2012JCP,Norizoe:2005,Griffiths:2011}. Polymer brushes also play a vital role in industrial applications, \textit{e.g.} surface coating, wetting, and colloidal stabilization~\cite{Kobayashi:2013}. Diblock-copolymer brushes and multicomponent-homopolymer brushes, \textit{i.e.} sophisticated brushes, have been extensively studied in recent years~\cite{O'Driscoll:2011,Guskova:2012,Wang:2009}. Polymers grafted onto spherical substrates or colloids have also drawn significant attention~\cite{Griffiths:2011,Norizoe:2005,Norizoe:2012JCP}. These complicated brushes show various phenomena according to the molecular architecture, mixing ratio between the polymer species, curvature and shape of the substrate, and other unique characteristics of each system. However, unique phase behavior of each system originated from such unique characteristics obstructs studies on the universal phase behavior of polymer brushes.

Therefore, the most basic and simplest polymer brush, \textit{i.e.} single-component linear homopolymers homogeneously grafted onto a planar substrate, was simulated in 3 dimensions (3-D) using the molecular Monte Carlo method to determine the universal and structural phase behavior of polymer brushes~\cite{Norizoe:2013EPL}. The simulation results showed that 2-dimensional (2-D) microphase separation of the grafted polymers in the lateral direction of the substrate occurs at low temperature. Various lateral domain patterns to minimize the interfacial free energy between low and high-density domains of the polymers are revealed. These domain patterns are similar to those found in diblock copolymer melts at microphase separation. These results demonstrate that topological constraints, \textit{i.e.} the grafting points of the brushes and connecting points between blocks of each block copolymer, universally yield the microphase separation and various domain patterns, independently of the unique characteristics of each system.

Early researchers also studied the single-component linear homopolymers grafted onto planar substrates and found one domain pattern; small circular clusters of the grafted polymers distributed over the substrate~\cite{Soga:1995,Lai:1992,Tang:1994,Grest:1993,Zhulina:1995,Yeung:1993,Koutsos:1997}. However, determination of the spatial arrangement of these circular clusters was reserved for future study. Furthermore, these early researchers missed the possibility of the other domain patterns.

Another universal structural phase behavior, \textit{i.e.} percolation and concomitant critical phenomena, has never been investigated in the homopolymer brush. Percolation transition is related to various physical phenomena and industrial applications, such as forest fire, diffusion in disordered media, string-like colloical assembly, and nano-switching devices~\cite{Stauffer1985,Norizoe:2005,Norizoe:2012JCP}. Furthermore, according to the general percolation theory~\cite{Stauffer1985}, critical phenomena, such as fractal structure of the percolation cluster and power law of cluster size distribution, are observed at the transition. We demonstrate that the percolation and critical phenomena also occur in brushes in the lateral direction of the substrate. The fractal structure of the percolation cluster of the grafted polymers is revealed and analyzed for the first time. The lateral structure of the brush depends on these phenomena as well as the 2-D microphase separation. These are also important for application, such as surface patterning, friction, and 2-D nano-switching devices.

Giving thought experiments, here we qualitatively discuss the percolation transition of the brush. Mushrooms of the grafted homopolymers are isolated from each other at extremely low $\rho$, where $\rho$ denotes the grafting density. In contrast, a polymer layer continuous in the lateral direction of the substrate covers the whole substrate at extremely high $\rho$. This indicates that, when $\rho$ rises from the extremely low value, percolation clusters of the grafted polymers, which bridge both the edges of the substrate, appear at an intermediate value of $\rho$. This percolation threshold (transition density) depends on the temperature, denoted by $T$, and the quality of the solvent because the size of the mushroom is dependent on these parameters. Therefore, a percolation transition line is constructed in $\rho T$-plane, in which typical phase diagrams of brushes are constructed for the sake of studying effects of the solvent quality and grafting density on the phase behavior~\cite{Norizoe:2013EPL}. Simulating the homopolymer brush in a coarse-grained scale using the molecular Monte Carlo method, we study these percolation and critical phenomena in the present work. The transition line, values of critical exponents, and fractal dimension of the percolation clusters are determined. We also discuss the relation between the percolation transition and microphase separation of the brush.

\section{Solvent-free model}
\label{sec:Solvent-freeModel}
In the present work, we employ a 3-D solvent-free coarse-grained model~\cite{Drouffe:1991,Daoulas:2010} of single-component linear homopolymers proposed by M{\"{u}}ller and Daoulas~\cite{DoctoralThesis,Norizoe:2010Faraday}. This was also utilized in our recent work~\cite{Norizoe:2013EPL} and quickly summarized in the present section.

The solvent-free model explicitly integrates out degrees of freedom of solvents, which are replaced with an effective non-bonded potential between solutes. This drastically diminishes degrees of freedom of the system and results in significantly reduced computational time required for simulation of the polymer brush laid in 3-D.

Harmonic spring potential, denoted by $H_\text{spr} = ( k_\text{spr} / 2 ) l^2$, with the spring constant $k_\text{spr} = 3 ( N - 1 ) / R_e^2$ linearly connects the coarse-grained segments in each polymer, where $l$ denotes the distance between the centers of the pair of the connected segments and $N$ is the number of segments per polymer. $R_e$ denotes the root mean square of the end to end distance of an ideal chain with the same molecular architecture. This $R_e$ is chosen as the unit length. This spring potential corresponds to the bonded interaction potential between the segments.

The free energy of non-bonded interactions in the solvent-free model is given by a functional, $H_{\text{non-bonded}}$, of the local segment density. We employ a third-order expansion of the non-bonded interaction free energy in a form of powers of the local segment density:
\begin{gather}
\label{eq:SingleComponentNonidealFreeEnergyDimensionless}
	\frac{ H_{\text{non-bonded}} }{k_B T} = \int_V \frac{dV}{{R_e}^3} \left( -\frac{1}{2}v' \left( \rho_p' ( \rVector ) \right)^2 + \frac{1}{3}w' \left(\rho_p' ( \rVector ) \right)^3 \right),  \\
	\rho_p' ( \rVector ) = \rho_s ( \rVector ) R_e^3 / N,  \notag
\end{gather}
where $k_B$ is the Boltzmann constant, $\rho_s ( \rVector )$ denotes the local volumetric number density of the segments at the spatial coordinate $\rVector$. The positive dimensionless constants, $v'$ and $w'$, set the attractive and repulsive interaction strengths among the segments, respectively. $\rho_p' ( \rVector )$ is the dimensionless local polymer density.

\begin{figure}[!tb]
	\centering
	\includegraphics[clip]{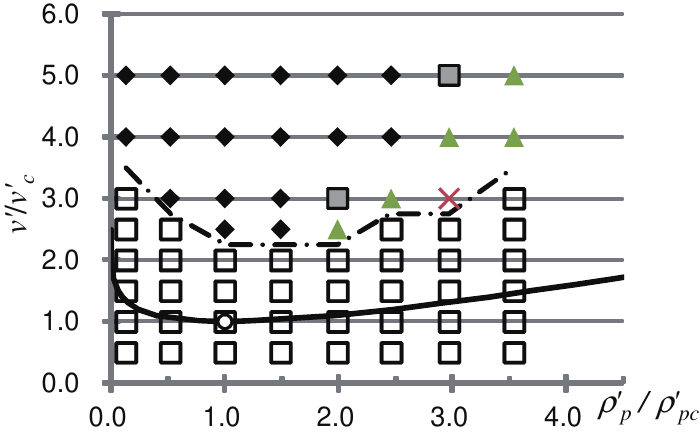}
	\caption{Phase diagram of the solvent-free system at fixed $w' = 0.01$. A solid black line denotes the binodal line of the system composed of the non-grafted polymers, calculated based on the mean-field theory. The macrophase separation of the non-grafted polymers occurs in regions of $v' / v'_c$ above this line, whereas the system is in the homogeneous phase below this line. The critical point, $( \rho_{pc}' , v'_c )$, is represented by a circle. On the other hand, a chain line denotes the binodal line of the polymer brush at $( L_x = 24.0 R_e, L_y = 24.0 R_e, L_z = 6.0 R_e )$. Open squares represent the homogeneous phase of the brush. Diamonds, triangles, and crosses represent the hexagonal, lamella, and inverse hexagonal structures of the brush~\cite{Norizoe:2013EPL}, respectively. The hexagonal and lamella structures are simultaneously observed in the system at the points of the grey squares.}
	\label{fig:BrushIniSqLatN32W001Lx240Ly240Lz60-PhaseDiagram}
\end{figure}

First, the phase behavior of a bulk solution of the non-grafted polymers modelled using the solvent-free system is discussed. This underlies the phase behavior of the brush of our model system~\cite{Norizoe:2013EPL}. The dimensionless average volumetric polymer density of the solution is denoted by $\rho_p' = n_p R_e^3 / V$, where $n_p$ and $V$ are the number of polymers and the system volume, respectively. At extremely large $v'$ and finite $w'$, macrophase separation of the polymers occurs, which results in a dense droplet floating in a dilute gas. In contrast, at finite $v'$ and extremely high $w'$, the system is in a homogeneous phase. The binodal line in $\rho_p' v'$-plane at fixed $w'$ is determined using mean-field approximation~\cite{DoctoralThesis,Doi:IntroductionToPolymerPhysics}. The critical point of this binodal line is denoted by $\rho_p' = \rho_{pc}'$ and $v' = v'_c$ (Ref.~\cite{DoctoralThesis}): $\left( \rho_{pc}' = 1 / \sqrt{ 2w' } \, , \, v'_c = 2 \sqrt{2w'} \, \right)$. In the present work, $w' = 0.01$ is fixed. This value of $w'$ determines $\rho_{pc}' = 7.071$, which was $w' = 0.0001$ and $\rho_{pc}' = 70.71$ in our previous work~\cite{Norizoe:2013EPL}. The present value of $\rho_{pc}'$, reduced from the previous work, decreases $n_p$ in a region of $\rho_p' / \rho_{pc}' \sim 1$, in which the simulation is performed. Therefore, with the present value of $w' = 0.01$, the amount of computation is significantly reduced, compared with the previous value of $w'$. The binodal line, calculated based on the mean-field theory at this fixed $w' = 0.01$ in the present work, is shown with a solid black line in Fig.~\ref{fig:BrushIniSqLatN32W001Lx240Ly240Lz60-PhaseDiagram}. This binodal line illustrates that the parameter $v'$ corresponds to the inverse temperature~\cite{Norizoe:2013EPL}. In other words, the quality of the implicit solvents at fixed $w'$ decreases with $v'$. Based on this phase diagram of the non-grafted polymers, the polymer brush is simulated and the structural phase diagram and percolation transition line are constructed at $w' = 0.01$.

\begin{figure*}[!tb]
  \centering
  \includegraphics[clip]{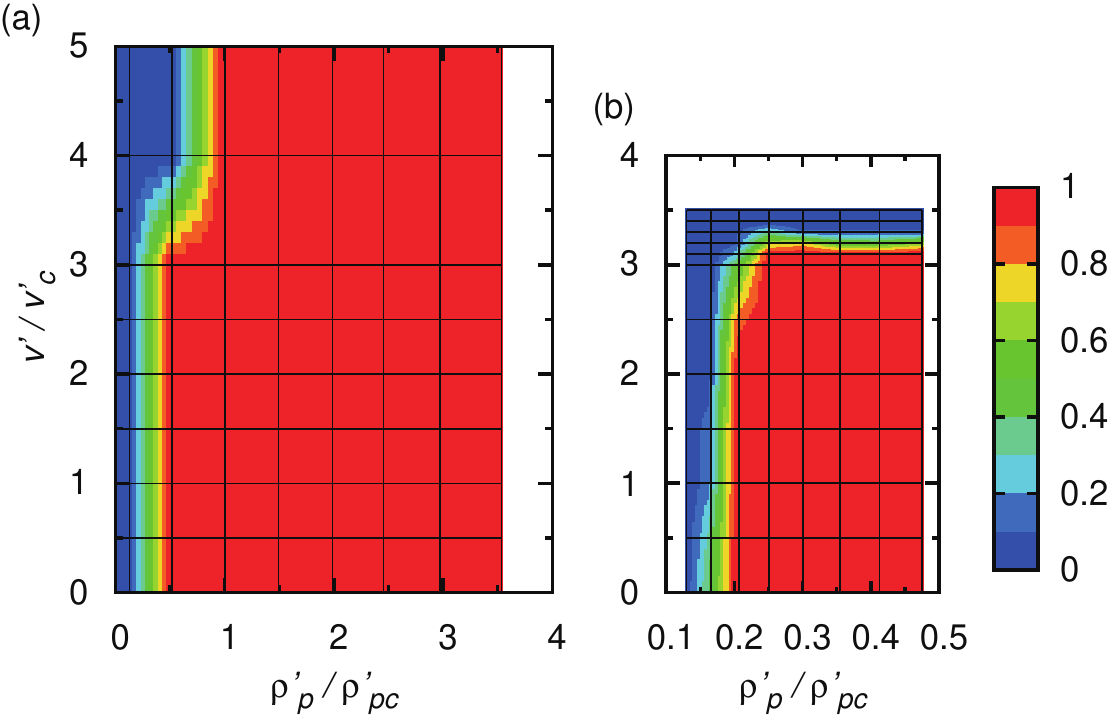}
  \caption{Occurrence probability of percolation clusters of the brush at $( L_x = 24.0 R_e, L_y = 24.0 R_e, L_z = 6.0 R_e )$. (b) The same graph as (a), constructed around a sharp boundary of (a) with a fine resolution. Grids denote the points where the simulation is performed, \textit{i.e.} the resolution of each panel. Note that panel (a) is constructed independently of the fine data of panel (b).}
  \label{fig:BrushIniSqLatN32W001Lx240Ly240Lz60AndFine3-PercolationForZeroDenTotal}
\end{figure*}
\begin{figure}[!tb]
	\centering
	\includegraphics[clip]{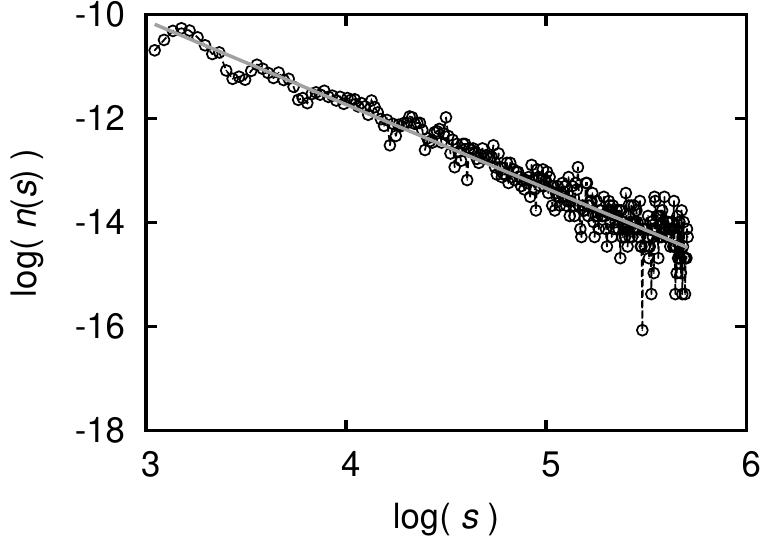}
	\caption{Cluster size distribution at $\rho_p' / \rho_{pc}' = 0.166$, $v' / v_c' = 0.01$, and $( L_x = 72.0 R_e, L_y = 72.0 R_e, L_z = 6.0 R_e )$. This parameter set is located on the percolation transition line in Fig.~\ref{fig:BrushIniSqLatN32W001Lx240Ly240Lz60AndFine3-PercolationForZeroDenTotal}. $\log (s)$-$\log (n(s))$ graph is plotted. A solid grey line shows a linear fitting of this graph, $y = -\tau * x + b$ with $\tau = 1.61$ and $b = -5.27$.}
	\label{fig:BrushIniSqLatN32W001Lx720Ly720Lz60RRc0166VVc001-ClusterSizeDistributionForZeroDen}
\end{figure}

\section{Simulation methods}
\label{sec:SimulationMethods}
Molecular Monte Carlo simulations are performed with the canonical ensemble in 3-D, using the standard Metropolis algorithm~\cite{ComputerSimulationOfLiquids,Frenkel:UnderstandingMolecularSimulation2002} at the fixed $w' = 0.01$. The Mersenne Twister algorithm is chosen as a random number generator~\cite{MersenneTwister1,MersenneTwister2,MersenneTwister3}. Thermal energy, $k_B T$, is chosen as the unit energy. $(L_x, L_y, L_z)$ denotes the size of the rectangular parallelepiped system box, where a periodic boundary condition is applied. This system box is laid in spatial regions of $0 \le \alpha < L_\alpha$, where $\alpha$ denotes the Cartesian direction. $N = 32$ and $\varDelta L = (1/6) R_e$ are fixed, where $\varDelta L$ defines the spatial range of the non-bonded interaction between the segments~\cite{Norizoe:2013EPL}. In one simulation step, a particle (segment) is randomly selected and given a uniform random trial displacement within a cube of edge length $2 \varDelta L$. One Monte Carlo step (MCS) is defined as $n_p N$ simulation steps, during which each particle is selected for the trial displacement once on average.
After $5.0 \times 10^5$ MCS, by which the system relaxes to the equilibrium state except at high $v' / v'_c$, we collect data every $10^4$ MCS till $10^6$ MCS and obtain 51 independent samples of particle configurations.

A terminal segment of each polymer is grafted onto a square lattice on a thin hard planar substrate laid at $z = 0$. This means the grafting points homogeneously distributed over the $xy$-plane of the system box. This substrate disallows the polymers to pass through and applies no other interactions, such as friction, to the polymers. The other segments of the grafted polymers are built in $z > 0$ and form a polymer layer in an interval of $0 < z \lesssim R_e$. The average volumetric polymer density in this region of the layer, defined as $\rho_p = n_p / ( L_x L_y \times R_e )$, corresponds to the average volumetric polymer density in the system of the non-grafted polymers. The parameter space of $( \rho_p' = \rho_p R_e^3, v' )$ of the brush is extensively searched in the simulation, and the structural phase diagram and percolation transition line are constructed in the $\rho_p' v'$-plane. We have verified that these phase diagram and transition line for the brush are not significantly changed when the grafting points of the polymers are randomly distributed over the substrate using uniform random numbers, which results in an inhomogeneous distribution of the grafting points.

Simulation results at $( L_x = 24.0 R_e, L_y = 24.0 R_e, L_z = 6.0 R_e )$ are mostly shown in the present work. We have verified, however, that the physical properties of the simulation system are not significantly changed even when the simulation is performed for a larger system with $( L_x = 72.0 R_e, L_y = 72.0 R_e, L_z = 6.0 R_e )$.

In the initial state, the conformation of each polymer is, using normal random numbers, arranged in random coils~\cite{Kawakatsu:StatisticalPhysicsOfPolymersAnIntroduction} with $\approx \! \! R_e$ of the root mean square end-to-end distance.

\section{Simulation results}
\label{sec:SimulationResults}
Here, the simulation results of the brush are discussed. The structural phase diagram of the brush is traced on the phase diagram of the non-grafted polymers given in Fig.~\ref{fig:BrushIniSqLatN32W001Lx240Ly240Lz60-PhaseDiagram}. In regions of $v' / v'_c$ below a chain line drawn in this phase diagram of the $\rho_p' v'$-plane, the homogeneous phase of the brush is observed. On the other hand, above the chain line, the 2-D microphase separation occurs. These results indicate this chain line denotes the binodal line of the brush. Next, we define the percolation cluster and discuss the percolation and critical phenomena of the brush.
\begin{figure*}[!tb]
  \centering
  \includegraphics[clip,width=12cm]{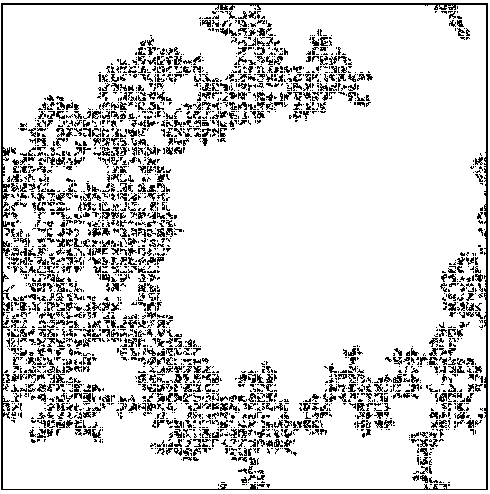}
  \caption{An example of a snapshot of the percolation cluster with the system size $( L_x = 72.0 R_e, L_y = 72.0 R_e, L_z = 6.0 R_e )$ at $\rho_p' / \rho_{pc}' = 0.166$ and $v' / v_c' = 0.01$, at which the percolation transition line runs. Sampled at 9 $\times 10^5$ MCS.}
  \label{fig:BrushIniSqLatN32W001Lx720Ly720Lz60RRc0166VVc001_000900000MCS-ConfPercolatedForZeroDen}
\end{figure*}
\begin{figure}[!tb]
	\centering
	\includegraphics[clip]{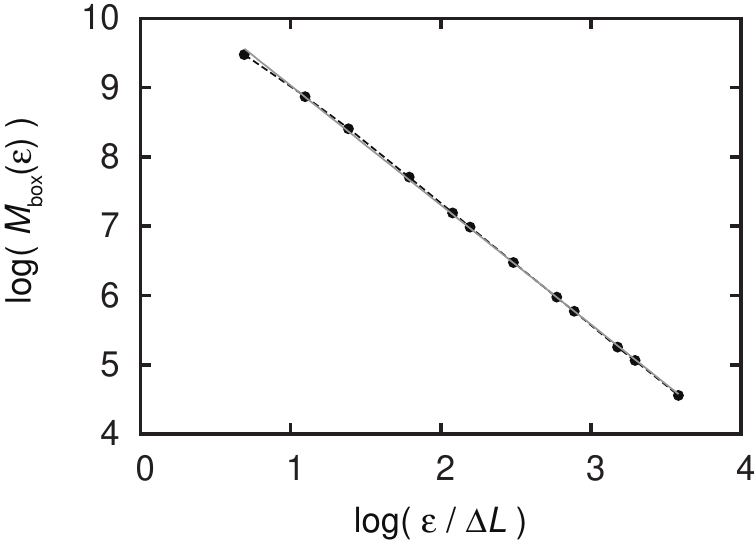}
	\caption{A result of box counting of the percolation clusters with the system size $( L_x = 72.0 R_e, L_y = 72.0 R_e, L_z = 6.0 R_e )$ at $\rho_p' / \rho_{pc}' = 0.166$ and $v' / v_c' = 0.01$, at which the percolation transition line runs. $\log (\epsilon)$-$\log (M_\text{box}(\epsilon))$ graph is plotted, where $\epsilon$ denotes the box size and $M_\text{box}(\epsilon)$ is the number of covered boxes with the size $\epsilon$ found in the system. A solid grey line shows a linear fitting of the graph, which is $y = -D_f * x + b$ with $D_f = 1.7$ and $b = 10.75$. $D_f$ is referred to as fractal dimension or box counting dimension.}
	\label{fig:BrushIniSqLatN32W001Lx720Ly720Lz60RRc0166VVc001-BoxCountForZeroDen}
\end{figure}

\subsection{Definitions of cluster size and percolation cluster}
\label{subsec:DefinitionsOfClusterSizeAndPercolationCluster}
First, a square lattice with the grid spacing $\varDelta L$ is set on the substrate. The index of the resulting ``pixel'' is denoted by $( i_x, i_y )$ for $0 \le i_\alpha < L_\alpha / \varDelta L$. We assume that a pixel $( i_x, i_y )$ is covered by the polymers when one or more segments are found in a square prismatic region ranging in $i_x \le x / \varDelta L < i_x + 1$,  $i_y \le y / \varDelta L < i_y + 1$, and $0 \le z < L_z$. A ``link'' is defined as adjacency between a pair of covered pixels satisfying a relation: $\left| i_x - i'_x \right| \le 1$ and $\left| i_y - i'_y \right| \le 1$, where $( i_x, i_y )$ and $( i'_x, i'_y )$ denote the indices of the pair respectively. A single network of links consisting of $M$ covered pixels is defined as a ``cluster with the size $M$''. A percolation cluster is defined as a large cluster that bridges both the edges of the substrate.

Linearly-connected segments of each grafted polymer range in a region of $0 \le z \lessapprox R_e = 6 \varDelta L$. The spatial range of the link, which is $\approx 3 \varDelta L$, covers almost a half or more of the thickness of this polymer layer when the cluster structure is also considered in $z$-direction. Furthermore, the linearly-connected segments connect the top of the polymer layer to the bottom through the intermediate. Therefore, even if the definition of the 3-D cluster structure is chosen for the analysis of the percolation phenomena, the percolation phenomena in the lateral direction of the substrate insignificantly depends on the cluster structure in $z$-direction.

\subsection{Percolation and critical phenomena}
\label{subsec:PercolationAndCriticalPhenomena}
Figure~\ref{fig:BrushIniSqLatN32W001Lx240Ly240Lz60AndFine3-PercolationForZeroDenTotal} shows occurrence probability of percolation clusters of the brush at each value set of $( \rho_p', v' )$. A sharp boundary divides percolation and non-percolation phases. The occurrence probability abruptly changes from 0 to 1 at this sharp boundary when $\rho_p'$ rises from extremely low values. This result is similar to the abrupt change of the occurrence probability at the percolation threshold in usual percolation phenomena~\cite{Stauffer1985,Norizoe:2005,Norizoe:2012JCP}. The cluster size distribution, denoted by $n(s)$ where $s$ is the cluster size, is calculated at $( L_x = 72.0 R_e, L_y = 72.0 R_e, L_z = 6.0 R_e )$ along the boundary curve of the occurrence probability of the percolation clusters. Here, the cluster size distribution is defined as, $n(s) = M(s) / N_\text{pixel}$, where $M(s)$ is the number of clusters with the size $s$ found in the system and $N_\text{pixel} = L_x L_y / ( \varDelta L )^2$ denotes the total number of the pixels. An example of $n(s)$ on the boundary curve is presented in Fig.~\ref{fig:BrushIniSqLatN32W001Lx720Ly720Lz60RRc0166VVc001-ClusterSizeDistributionForZeroDen} in double logarithmic scales, which shows a relation, $n(s) \propto s^{-\tau}$ with $\tau = 1.6$. This power law with $\tau = 1.6$ is unchanged along the boundary curve. The power law of the cluster size distribution is also observed in usual percolation phenomena at the percolation threshold~\cite{Stauffer1985,Norizoe:2005,Norizoe:2012JCP}. These results indicate that the sharp boundary is the percolation transition line of the grafted polymers. The percolation transition occurs in regions of $\rho_p' = \rho_p R_e^3 \sim 1$, since the mushrooms, whose size is $\approx \! \! R_e$, start to overlap with each other at this $\rho_p'$ when $\rho_p'$ is raised from extremely low values. $\tau$ is referred to as Fisher exponent, a critical exponent, of the transition of percolation systems. This $\tau = 1.6$ is slightly smaller than the results of a model system of dispersion of polymer-grafted colloids, $\tau = 1.9$ in 2-D and 2.2 in 3-D~\cite{Norizoe:2005,Norizoe:2012JCP}. Other critical exponents are not calculated in the present work because of poor resolution of the data.

Here, the relation between the percolation transition and 2-D microphase separation of the brush is discussed. When the microphase separation occurs, the mushrooms are merged into high-density domains. This decreases the effective grafting density of the polymers and results in an abrupt rise of the percolation threshold at the outset of the microphase separation, \textit{i.e.} around the region where the percolation transition line and binodal line of the brush cross. The percolation transition and binodal lines of the brush in regions of $0.129 \lessapprox \rho_p' / \rho_{pc}' \lessapprox 1.00$ and $3.0 \lessapprox v' / v_c' \lessapprox 4.0$ in Figs.~\ref{fig:BrushIniSqLatN32W001Lx240Ly240Lz60-PhaseDiagram} and \ref{fig:BrushIniSqLatN32W001Lx240Ly240Lz60AndFine3-PercolationForZeroDenTotal} illustrate these results.

Finally, we study structure of the percolation cluster at the percolation transition. As an example, a snapshot of the percolation cluster at $\rho_p' / \rho_{pc}' = 0.166$ and $v' / v_c' = 0.01$, at which the percolation transition line runs, is presented in Fig.~\ref{fig:BrushIniSqLatN32W001Lx720Ly720Lz60RRc0166VVc001_000900000MCS-ConfPercolatedForZeroDen}. A complex pattern is found in this snapshot, as expected. At this value set of $\rho_p' / \rho_{pc}'$ and $v' / v_c'$, the fractal dimension of the percolation clusters, denoted by $D_f$, is measured using box counting method. The result of the box counting is given in Fig.~\ref{fig:BrushIniSqLatN32W001Lx720Ly720Lz60RRc0166VVc001-BoxCountForZeroDen}. This indicates $D_f = 1.7$, which is unchanged along the percolation transition line. The unchanged value of $D_f$ is also found in usual percolation phenomena~\cite{Stauffer1985}. $D_f = 1.7$ is qualitatively consistent with the result of conventional 2-D lattice models of percolation, $D_f = 91 / 48 = 1.9$ (Ref.~\cite{Stauffer1985}).

In the present simulation, the polymers are grafted onto a square lattice. This corresponds to a conventional 2-D square lattice model of percolation at $p = 1$, where $p$ denotes the probability of a site being occupied by a particle. This conventional system is always in the percolation phase at $p = 1$. However, in the case of the brush, both of the size of the mushrooms, which corresponds to the particle diamter of the lattice model, and the grafting density, which corresponds to the lattice constant, can be changed. Furthermore, the center of mass of the mushroom moves around and is displaced from the grafting points, which results in inhomogeneity of the lattice. The effective grafting density decreases when the microphase separation occurs, as discussed. These features of the brush change the effective value of $p$ and both the percolation and non-percolation phases are observed in the present system, although the corresponding conventional lattice model is always in the percolation phase. This inhomogeneity of the lattice also indicates that, as discussed in section~\ref{sec:SimulationMethods}, the simulation results do not significantly change when the grafting points are randomly distributed over the substrate. This shows that our findings are also demonstrated in experiments, in which the polymers are randomly grafted.

\section{Conclusions}
\label{sec:Conclusions}
In conclusion, the percolation and critical phenomena of the homopolymer brush have been studied using the molecular Monte Carlo simulation in a coarse-grained scale in 3-D. The structural phase diagram and percolation transition line have been constructed. The relation between the percolation transition and binodal lines of the brush has been examined. Fisher exponent of the system, $\tau = 1.6$, has been measured. Fractal structure of the percolation clusters at the percolation transition line has been revealed. $D_f = 1.7$.

These results can also be confirmed in arbitrary polymer brushes. In other words, the percolation and critical phenomena are universally found in brushes. However, the values of $\tau$ and $D_f$ could depend on the characteristics of each system.

\begin{acknowledgments}
The authors wish to thank Prof. Marcus M{\"u}ller, Dr Kostas Ch. Daoulas, and Prof. Thein Kyu for helpful suggestions and discussions.
\end{acknowledgments}

%

\end{document}